\documentclass[review,authoryear]{elsarticle}
\usepackage{courier,url}
\usepackage{amsmath,amsfonts,amssymb}
\usepackage{amsthm}
\usepackage{natbib} 
\usepackage{graphicx} 
\graphicspath{{figures/}}
\usepackage{marginnote} 

\newcommand{\ie}{{\it i.e.},\ }   
\newcommand{\eg}{{\it e.g.},\ }   

\newcommand{\ud}{\, \mathrm{d}}

\DeclareMathOperator{\BigO}{O}

\long\def\comment#1{}

\usepackage{lineno}
\providecommand{\linenumbers}[1]{} 

\begin{document}
\title{Frequency responses of age-structured populations: Pacific salmon as an example}

\author[mcm,des]{Lee~Worden\corref{cor}}
\ead{worden.lee@gmail.com}
\author[wfb]{Louis~W.~Botsford}
\ead{lwbotsford@ucdavis.edu}
\author[des]{Alan~Hastings}
\ead{amhastings@ucdavis.edu}
\author[wfb]{Matthew~D.~Holland}
\ead{mdholland@ucdavis.edu}

\cortext[cor]{Corresponding author}

\address[mcm]{Department of Biology, Theoretical Biology Lab, McMaster University, LS 332 1280 Main Street West, Hamilton, ON L8S 4K1 Canada}
\address[des]{Department of Environmental Science \& Policy, University of California, Davis, One Shields Ave, Davis, California 95616 USA}
\address[wfb]{Department of Wildlife, Fish, \& Conservation Biology, University of California, Davis, One Shields Ave, Davis, California 95616 USA}

\begin{abstract}
Increasing evidence of the effects of changing climate on physical ocean conditions and long-term changes in fish populations adds to the need to understand the effects of stochastic forcing on marine populations. Cohort resonance is of particular interest because it involves selective sensitivity to specific time scales of environmental variability, including that of mean age of reproduction, and, more importantly, very low frequencies (i.e., trends). We present an age-structured model for two Pacific salmon species with environmental variability in survival rate and in individual growth rate, hence spawning age distribution. We use computed frequency response curves and analysis of the linearized dynamics to obtain two main results. First, the frequency response of the population is affected by the life history stage at which variability affects the population; varying growth rate tends to excite periodic resonance in age structure, while varying survival tends to excite low-frequency fluctuation with more effect on total population size. Second, decreasing adult survival strengthens the cohort resonance effect at all frequencies, a finding that addresses the question of how fishing and climate change will interact.
\end{abstract}


\maketitle

\linenumbers

\section{Introduction}

In population theory, interest is increasing in the complex ways in which age-structured,
density-dependent populations respond selectively to different time
scales (or frequencies) of variability in the environment.  These questions dovetail with the increasing practical
need to understand how populations will respond to potential changes
in life history parameters and the time scales of environmental
variability due to climate change and increasing pressure on natural
resources such as fisheries.  There is a growing awareness that model
population responses can appear to filter certain frequencies in
environmental variability \citep{greenman:2005:freq}.  Furthermore, these
effects can serve to amplify variability by exciting modes of population behavior
that without environmental variability would be locally stable
\citep{greenman:2003:ampli}.  One example of such behavior, cohort
resonance, is closely identified with life history characteristics in
that it features cycles of period equal to the dominant age of reproduction.
\citep{bjornstad:1999:cycl}.

There is particular interest in the dynamic responses of the marine fish targeted by global fisheries
as they are subject to a dramatic artificial change in a life history
parameter, adult survival, and there are indications that the dominant
time scales of environmental variables such as ENSO that affect fish populations may
change with a changing climate \citep{Timmermann:1999:incr}.
Empirical observations and models indicate that 
variability in fish populations increases with increased fishing
\citep{hsieh:2006:fish,anderson:2008:whyf}.  The
cohort resonance phenomenon in fish populations presents particular
problems for climate change because it enhances sensitivity to very
slow signals (trends) in addition to those at the period of dominant age of
reproduction \citep{bjornstad:2004:tren}.  Observations of such effects could be
confounded with potential slow changes in the environment, and thus
make it difficult to differentiate between them.

Within the general concern for the combined effects of fishing and climate change
on marine ecosystems \citep{perry:2010:sens}, there is a particular interest in the effects of the
marine environment and fishing on population dynamics of Pacific salmon on annual
to decadal time scales. Analyses of the influences of the ocean
environment on Pacific salmon, whether statistical examination of
covariability between population and environmental variables
\citep[\eg][]{logerwell:2003:trac} or estimation of survivals to
specific sizes and ages through analysis of coded wire tagging data
\citep[\eg][]{coronado:1998:spat,teo:2009:spat}, commonly assume the
variable ocean environment influences survival during the early ocean
phase of this anadromous genus.  In fewer cases 
environmental variability in the age of maturation has been the dependent population variable
\citep[\eg][]{pyper:1999:patt}.

These empirical findings raise questions regarding the relative roles
of random variability in survival at various ages, and in the age
distribution of reproduction, in salmon population dynamics. Random
survival is typically assumed to influence abundance directly, but the
effects of varying age of spawning are not as clear, and there is a
need to understand differences in population response to these two
sources of random variability.

Here we examine and compare the effects of environmental variability
in survival and age of spawning on the magnitude and time scales of
population variability.  We are engaged in a study of the influence of
the oceanographic environment on Pacific salmon, the GLOBEC North East
Pacific program, part of the US Climate Change program.  Earlier
retrospective analyses indicated differences in the responses of the
two California Current congeners, chinook salmon, {\it Oncorhynchus
tshawytscha}, and coho salmon, {\it O. kisutch}, and hence we are
interested in the population dynamic differences between them
\citep{botsford:2002:patt}.  For instance, coho salmon collapsed synchronously along the coast between 1980 and 2000, and coho appear to have more high frequency variability than chinook in their catch record.  These species differ in their age
distribution of spawning and in other ways \citep{botsford:2005:diff}.
We have explored some of the differences in probabilities of
extinction of populations with these spawning age distributions in
response to time-varying marine survival
\citep{hill:2003:effe,botsford:2005:diff}.  In salmon, differences in
maturity schedule are due to differences in size distribution (and
therefore prior growth rate)
\citep{young:1999:envi,vollestad:2004:effe}.  Similar effects occur in
other non-salmonid species \citep{day:2002:deve}.  To avoid confusion,
we note that the mechanisms studied here are neither (1) the indirect
effect on survival of varying development rate due to consequent
variation in time spent at higher mortality
\citep[\eg][]{moloney:1994:deve}, nor (2) the interaction between age
structure and over-compensatory density-dependent recruitment that
causes cycles of period roughly twice the generation time
\citep{ricker:1954:stoc,botsford:1997:dyna}.

In the models used here oscillations with period $2T$ do not
arise (where $T$ is generation
time or age of dominant effect on recruitment), primarily because of
the form of stock-recruitment function we use, but
the cyclic mode with period $T$ turns out to play a central role in our
investigation.  Recently \citet{myers:1998:simp} explored stochastic
forcing of a cyclic mode of variability of period $T$ as a potential
cause of observed cycles in sockeye salmon ({\it Oncorhynchus nerka})
in British Columbia, Canada.  This mode is similar to the ``echo effect''
associated with linear age-structured models of semelparous species
\citep{sykes:1969:ondi}.  Myers et al.\ noted that while this mode would not be
the dominant mode, it could appear clearly in solutions obtained
through forcing by randomly varying survival.  This was essentially cohort resonance, though they did not use that term.

The time scales (or frequency content) of the population response 
also depend on the mode of observation, \eg whether the data in a time
series are recruitment, abundance or catch
\citep{botsford:1986:effe,anderson:2008:whyf}. The nature of a catch
time series depends on fishing mortality rate, with higher fishing
rates leading to higher frequencies of variability
\citep{botsford:1986:effe,hsieh:2006:fish}.  Differences between the
spectra of recruitment and abundance have been illustrated by
\citet{bjornstad:2004:tren} for Atlantic cod ({\it Gadus morhua}).  The most commonly available measurement
of salmon populations is annual catch, either in numbers or biomass,
and occasionally the age distribution of catch is determined.
Frequently catch can be assumed to be individuals who if not caught would be spawning
that year because they are near the mouth of (or in) the spawning
river.  In some streams, spawning escapement is also estimated, and
sometimes the age composition of spawners is also estimated.  Catch
and escapement can be summed to obtain the total abundance of potential spawners
in a year.  Total population abundance at time $t$ cannot be observed
directly, but estimates can be obtained through cohort reconstruction from
several years of age-specific data from catch and escapement.

Here we explore several aspects of the spectral response of
marine fish population dynamics, using parameter values representative of two species of Pacific salmon over a
range of survivals as examples.  Our practical interests are in the
combined effects of fishing and climate change, so we focus on 
effects of long-term changes in survival on the population response to various time scales
of environmental variability. Note, however, that long-term declines in survival can also be caused by the climate, independent of the fishery, as occurred in California Current coho salmon between 1980 and 2000 \citep{botsford:2005:diff}. We also explore the differences in responses
to environmental forcing at different points in the life history
(i.e. development rate vs. survival), and spectral
differences between various kinds of observations (e.g., observations of recruitment vs. observations of 
total abundance).  We present results in terms of the changing
spectral responses, and relate the spectral response to the eigenvectors 
and eigenspace structure of the linearized model.

\section{Model Formulation}

To investigate these questions, we introduce a density-dependent,
stochastic age-structured model which we can use to compare the
effects of temporal variability in the timing of reproduction and annual survivorship of salmon. Let $n$ be
the maximum spawning age in years, and let
$\vec{x}(t)=(x_1(t),\dots,x_n(t))^{\mathrm{T}}$ be the age-structured
population vector at year $t$. Then our model is
\begin{equation} \label{eq:pop-dyn}
\vec{x}(t) = F(\vec{x}(t-1),t) = \left( \begin{array}{c} 
R(P(t)) \\ 
s(t)\,x_1(t-1) \\
s(t)\,x_2(t-1) \\
\vdots \\
s(t)\,x_{n-3}(t-1) \\
\big(1-\delta_e(t) \big)\,s(t)\,x_{n-2}(t-1) \\
\delta_l(t)\,s(t)\,x_{n-1}(t-1) \end{array} \right)
\end{equation}
where
$P(t)=(\delta_e(t)\,x_{n-2}(t-1)+(1-\delta_l(t))\,x_{n-1}(t-1)+x_n(t-1))$
is the number of fish returning to spawn in year $t$, and
\begin{equation}
  \label{eq:beverton-holt}
  R(P(t))=\frac{\alpha P(t)}{1+\beta P(t)}
\end{equation}
is a Beverton-Holt density dependent recruitment term
\citep{beverton:1957:dyna}. Thus the recruitment $R(P(t))$ represents the number of
outmigrants (smolts) leaving a spawning stream in year $t$, resulting from egg
production by individuals of various spawning ages.
We have assumed that migration from freshwater to the marine environment occurs in the first year, and that there is no difference in fecundity between the different ages of spawning.  Parameters $\alpha$ and $\beta$ characterize the density-dependent reproductive phase: $\alpha$ is the density-independent per-capita growth rate when the population is very small, and $\frac{\alpha}{\beta}$ is the maximum total number of offspring in the population. Timing of reproduction is controlled by the remaining two parameters, $\delta_e(t)$ and $\delta_l(t)$: most individuals spawn at age $n-1$; a proportion $\delta_e(t)$ of those surviving to age $n-2$ spawn at that age in year $t-1$; and a proportion $\delta_l(t)$ of age $n-1$ fish in year $t$ postpone spawning until year $t+1$, when they are age $n$. Annual survival $s(t)$, as written here, affects all age classes in year $t$. However, we also explore the possibility that the dominant variability in ocean survival occurs during the period immediately following ocean entry.  This is a challenging stage for juvenile salmon, as they are completing the transition from  freshwater to  the marine environment, and they are dependent on the highly variable food production in the coastal ocean.   

\begin{figure}
\begin{center}
\scalebox{1}{\includegraphics{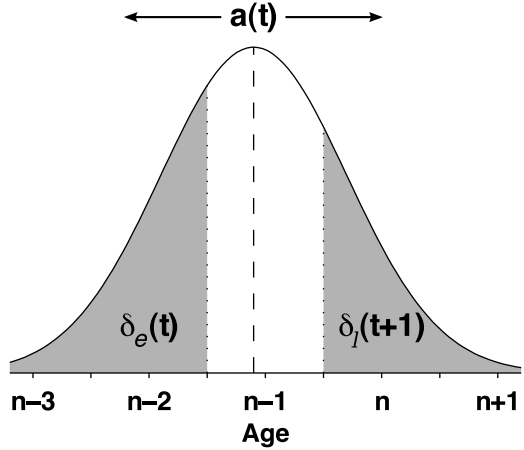} }
\end{center}
\caption{An example illustrating how the fraction of early and late spawners is modeled.  The central spawning age $a(t)$ varies from year to year.  Here we picture a normal distribution of potential spawning age with mean $a(t)=n-1.1$ and $\sigma=0.8$. The parameters representing the proportions of early spawners ($\delta_e$) in year $t$ and late spawners ($\delta_l$) in year $t+1$ are the integrals of this function over $(-\infty,n-1.5]$ and $[n-0.5,\infty)$, respectively.}
\label{fig:a-distr}
\end{figure}

We are interested in
understanding the time scales of variation in the solutions to the model
(equation~$\ref{eq:pop-dyn}$) in response to fluctuations in annual
survivorship and timing of spawning.  We do this by studying three
cases: the case of fluctuating survival $s(t)$ at all ages, the case
in which survival varies but only at the age of entry into the ocean,
and the case in which fluctuating mean age of spawning $a(t)$ produces
yearly changes in $\delta_e (t)$ and $\delta_l (t)$. In all cases, the 
environmental fluctuation is a Gaussian white-noise signal, $\xi(t)\in\mathbb{R}$ with $\mathrm{E}(\xi(t))=0$.  To model
fluctuating survival we use $s(t) = s^\circ + \xi(t)$, 
where $s^\circ$ is the unperturbed, constant survival.  In the simulation the distribution of $s(t)$ is truncated to ensure that $0\leq s(t)\leq1$.
  To model
fluctuating age of spawning, we suppose that individuals' ages of
spawning are chosen from a normal distribution whose mean is the
central age of spawning $a(t)$ in year $t$ (Fig.~\ref{fig:a-distr}),
specifically 
\begin{equation}
 p(a-a(t)) = \frac{1}{\sqrt{2 \pi}\sigma} e^{-\left(\frac{a-a(t)}{\sqrt{2}\sigma}\right)^2}.
\end{equation} 
\noindent
We approximate early and late spawning by having all early spawners
spawn at age $n-2$ and all late spawners spawn at age $n$, so
that 
\begin{equation}
 \delta_e(t)=\int^{n-1.5}_{-\infty} p(a-a(t))\ud{a}
\end{equation} and 
\begin{equation}
 \delta_l(t)=\int_{n-0.5}^{\infty} p(a-a(t-1)) \ud{a}.  
\end{equation}
From the point of view of a cohort, the portions of that cohort that
will spawn early and late, \ie $\delta_e$ and $\delta_l$, are set by
the value of $a(t)$ in the year that that cohort transitions from age
$n-2$ to age $n-1$.  That is, $\delta_e(t)$ is a function of the
central spawning age in year $t$, whereas $\delta_l(t)$, which has its
effect a year after it is determined, is a function of the central
spawning age in year $t-1$.  We model the fluctuating central age of
spawning as $a(t)=a^\circ+\xi(t)$.

To compare the dynamics of the two salmon species along the west coast of the contiguous U.S., we have used parameter values that approximate known or likely values for the populations of coho and chinook salmon.  Survival rates in the ocean are reasonably well known, and we consider three cases: a ``typical'' value of $0.85~\mathrm{yr}^{-1}$ \citep{bradford:1995:comp}, a ``small'' value of $0.28~\mathrm{yr}^{-1}$, and a ``very small'' value of $0.2~\mathrm{yr}^{-1}$, estimated from more recent observations of coho salmon.  Since the ``very small'' value is not sustainable in the chinook model, that case is not considered.  The distributions of values of $\alpha$ and $\beta$ (equation~$\ref{eq:beverton-holt}$) have been estimated for coho salmon, but not for chinook salmon \citep{barrowman:2003:vari}, and we use the modes from those distributions ($\alpha=60$ and $\beta=0.00017$). In chinook salmon populations in the California Current, individuals typically spawn primarily at a single dominant age, with less spawning at adjacent ages.  The dominant age of spawning increases with latitude \citep{hill:2003:effe}.  Here we chose dominant spawning at age 4, i.e., $n=5$ and $a^\circ=4$.  Precocious spawning in chinook salmon ranges from 0--10 percent in males and from 0--3 percent in females \citep{healey:1991:life}.  We use $\sigma=0.4$, which makes $\delta_e=\delta_l\approx0.1$.  Coho salmon in the California Current spawn predominantly at age 3 with substantial precocious spawning at age 2 and minimal spawning at age~4.  Precocious spawning in coho salmon consists almost completely of males, is more variable than in chinook salmon, and can be as high as 30 percent in natural wild populations \citep{sandercock:1991:life}.  It is not possible to observe the effects of precocious spawning by males on reproduction directly, but for coho salmon, a modification of genetic methods for estimating effective population size indicated the effective proportion of 2-year-olds to be 35 percent in two naturally spawning populations \citep{doornik:2002:patt}.  For coho salmon we chose $n=4$, $a^\circ=2.75$ and $\sigma=0.2$, which makes $\delta_e\approx0.1$ and $\delta_l\approx0.0001$.  The standard deviation of the stochastic variation,
$\sigma_E$, is $0.085$ for variation in survival, and $0.2$ for
variation in mean spawning age.
\begin{table}[htdp]
\caption{Model parameters and unperturbed values.}
\begin{center}
\begin{tabular}{|c|p{2in}|c|c|}
	\hline
	Parameter & Description & Coho Value & Chinook Value \\
	\hline
	$s^\circ$ & annual survival rate, all ages & 0.85, 0.28, 0.2 & 0.85, 0.28 \\
	\hline
	$\delta_e$ & proportion of age $n-2$ fish spawning & 0.1056 & 0.1056 \\
	\hline
	$\delta_l$ & proportion of age $n-1$ fish delaying spawning to age $n$ & $8.84\times 10^{-5}$ & 0.1056 \\
	\hline
	$\alpha$ & slope at origin of Beverton-Holt stock-recruitment curve & 60 & 60 \\
	\hline
	$\beta$ & saturation parameter of Beverton-Holt stock-recruitment curve & $1.7\times 10^{-4}$ & $1.7\times 10^{-4}$ \\
	\hline
	$a^\circ$ & central age of spawning & 2.75 & 4 \\
	\hline
	$\sigma$ & standard deviation of spawning age distribution & 0.2 & 0.4 \\
	\hline
	$\sigma_E$ & standard deviation of environmental forcing signal & 0.085 (survival) & 0.085 (survival) \\
	 &  & 0.2 (spawning age) & 0.2 (spawning age) \\
	\hline
\end{tabular}
\end{center}
\label{default}
\end{table}

\section{Model Analysis}

In each of the model cases, the age vector $\vec{x}$ converges to a
neighborhood of a single point, and fluctuates in that neighborhood in
response to the fluctuating life-history parameters.  To find the
center point, we consider the deterministic system defined by
equation~$\ref{eq:pop-dyn}$ with $s$, $\delta_e$, and $\delta_l$ all fixed at
their unperturbed values.  This system has a unique positive fixed
point that is globally attracting for positive trajectories:
\begin{equation}\label{eq:fixed-point} 
\tilde{\vec{x}} = \left(\begin{array}{c}
 1 \\ s \\ s^2 \\ \vdots \\ s^{n-3} \\ (1- \delta_e)s^{n-2} \\
 (1- \delta_e)\delta_l s^{n-1} \end{array}\right)
\frac{\alpha - c}{\beta}, 
\end{equation} 
where 
$1/c=\delta_e s^{n-3} + (1-\delta_e )(1-\delta_l )s^{n-2}+(1-\delta_e )\delta_l s^{n-1}$ 
is the amount of spawning in the lifetime of the average smolt.

As we will see, we can understand a great deal about the population
dynamics by looking at the linearized dynamics at this fixed point of
the deterministic dynamics.  We linearize the effect of the stochastic
noise term $\xi(t)$ as well as the age-structured population vector
$\vec{x}(t)$:
\begin{equation}
  \label{eq:linearization}
  \vec{y}(t) \approx \mathbf{J}\vec{y}(t-1)+\sum_{l=0}^L\vec{H}(l)\xi(t-l),
\end{equation}
where $\vec{y}(t)=\vec{x}(t)-\tilde{\vec{x}}(t)$ is the deviation from
the fixed point, $\mathbf{J}$ is the Jacobian matrix of the dynamics
at the fixed point, and linearizing in the noise term gives a sequence
of ``forcing vectors'' $\vec{H}$ that summarizes how the noise
enters into the dynamics with various time lags (see the Appendix for
details).

At the fixed point, the Jacobian matrix is
\begin{equation}
  \mathbf{J} =
   \left(\begin{array}{cccccccc}
    0 & \cdots & 0 & \delta_e \frac{c^2}{\alpha}
                                   & (1-\delta_l)\frac{c^2}{\alpha}
                                                & \frac{c^2}{\alpha} \\
    s &        &   &               &            & 0                  \\
      & \ddots &   &               &            &                    \\
      &        & s &               &            & \vdots             \\
      &        &   & (1-\delta_e)s &            &                    \\
      &        &   &               & \delta_l s &  0
   \end{array}\right),
\end{equation}
where all entries left blank are zero.

A linear system like this one decomposes naturally into independent
subsystems located in linearly independent one- and two-dimensional
subspaces, each characterized by a real eigenvalue or a complex pair
of eigenvalues \citep{Hirsch:1974:diff}.  The right eigenvectors of
the Jacobian matrix are the basis vectors for each of these subspaces
of the dynamics.  The complex conjugate eigenvalues of the Jacobian predict the resonant
frequencies of the population's response to noise.  In the
time-varying-survival cases, we only need a single forcing vector
$\vec{H}$, because the state of the environment in year $t$,
$\xi(t)$, only affects survival in year $t$; but in the
varying-age-structure case, because conditions in year $t$ affect the
number of early returns $\delta_e$ in year $t$ and the number of late
returns $\delta_l$ in year $t+1$, we have to include two forcing
vectors to describe the effects with and without one year's time lag.

In the case of fluctuating survival at all ages, the
forcing is captured by the vector $\vec{H}_s(0)$, with
\begin{equation} \label{eqn:Hs}
\vec{H}_s(0) = \left(\frac{\partial F_i}{\partial \xi(t)}\right)
    = \left( \begin{array}{c}
        0 \\ \tilde{x}_1 \\ \tilde{x}_2 \\ \tilde{x}_3 \\ \vdots \\
        \tilde{x}_{n-3} \\ (1-\delta_e)\tilde{x}_{n-2} \\ \delta_l \tilde{x}_{n-1}
      \end{array} \right)
    = \left( \begin{array}{c}
        0 \\ 1 \\ s^\circ \\ (s^\circ)^2 \\ \vdots \\
        (s^\circ)^{n-4} \\ (1-\delta_e)(s^\circ)^{n-3} \\ (1-\delta_e)\delta_l (s^\circ)^{n-2}
      \end{array} \right) \frac{\alpha-c}{\beta}.
\end{equation}
We note that when survival is forced additively, as
we have done here, the deterministic system described above is not
exactly the mean of the stochastic system. A stochastic system with
$s(t) = s^\circ e^{\xi(t)}$ would have the deterministic system as its
mean, but in the limit of small noise, these representations have
identically-shaped frequency responses that merely differ by a factor
of $s$. We chose additive noise for convenience.

For forcing of survival at ocean entry, only survival to age $2$ is
subject to fluctuation, so that
\begin{equation}  \label{eqn:Hse}
  \vec{H}_{s_e}(0) = \left( \begin{array}{c}
      0 \\ \frac{\alpha-c}{\beta} \\ 0 \\ \vdots \\ 0
      \end{array} \right).
\end{equation}

For time-varying ages of maturation, we have lag $0$ effects from
early spawning
\begin{equation} \label{eqn:Ha0}
\vec{H}_a(0) = \left(\frac{\partial F_i}{\partial \xi(t)}\right) =
\left(\frac{\partial F_i}{\partial \delta_e} \frac{\partial \delta_e}{\partial \xi(t)}\right)
\end{equation}
and lag $1$ effects from late spawning
\begin{equation} \label{eqn:Ha1}
\vec{H}_a(1) = \left(\frac{\partial F_i}{\partial \xi(t-1)}\right) =
\left(\frac{\partial F}{\partial \delta_l} \frac{\partial \delta_l}{\partial \xi(t-1)}\right).
\end{equation}
Since $\frac{\partial a(t)}{\partial \xi(t)}=1$ we have
\begin{equation} \label{eqn:dde}
\frac{\partial \delta_e(t)}{\partial \xi(t)} = \int^{n-1.5}_{-\infty}
-p'(a-a(t))\ud a = -p(n-1.5-a(t))
\end{equation}
and
\begin{equation} \label{eqn:ddl}
\frac{\partial \delta_l(t)}{\partial \xi(t-1)} = \int_{n-0.5}^{\infty}
-p'(a-a(t-1))\ud a = p(n-0.5-a(t-1)).
\end{equation}
The vector derivatives we need are
\begin{equation} \label{eqn:dFdde}
\left(\frac{\partial F_i}{\partial \delta_e}\right)
= \left( \begin{array}{c} s^{n-3}\frac{c^2}{\alpha} \\ 0 \\ \vdots
\\ 0 \\ -s^{n-2} \\ 0 \end{array} \right) \frac{\alpha -c}{\beta}
\end{equation}
and
\begin{equation} \label{eqn:dFddl}
\left(\frac{\partial F_i}{\partial \delta_l}\right)
= \left( \begin{array}{c} -(1-\delta_e^\circ)s^{n-2}\frac{c^2}{\alpha} \\ 0 \\ \vdots \\ 0 \\ (1-\delta_e^\circ)s^{n-1} \end{array} \right)
\frac{\alpha -c}{\beta},
\end{equation}
where $\delta_e^\circ$ refers to the unperturbed
value of $\delta_e$.  Substituting (\ref{eqn:dde}) and
(\ref{eqn:dFdde}) into (\ref{eqn:Ha0}) yields the forcing at lag $0$,
\begin{equation}
\vec{H}_a(0) = \left( \begin{array}{c} 
    -p(n-1.5-a^\circ) s^{n-3}\frac{c^2}{\alpha} \\
    0 \\ \vdots \\ 0 \\ 
    p(n-1.5-a^\circ)s^{n-2} \\
    0 
  \end{array} \right)
\frac{\alpha-c}{\beta}.
\end{equation}
Substituting (\ref{eqn:ddl}) and (\ref{eqn:dFddl}) into (\ref{eqn:Ha1}) 
gives the forcing at lag $1$,
\begin{equation}
\vec{H}_a(1) = \left( \begin{array}{c} 
 - p(n-0.5-a^\circ) (1-\delta_e^\circ)s^{n-2}\frac{c^2}{\alpha} \\
 0 \\ \vdots \\ 0 \\ 
p(n-0.5-a^\circ)(1-\delta_e^\circ)s^{n-1} \end{array} \right)
\frac{\alpha-c}{\beta}.
\end{equation}

\section{Results}

\subsection{Dynamics near the fixed point}

\begin{figure}
\begin{center}
\includegraphics[width=0.9\textwidth]{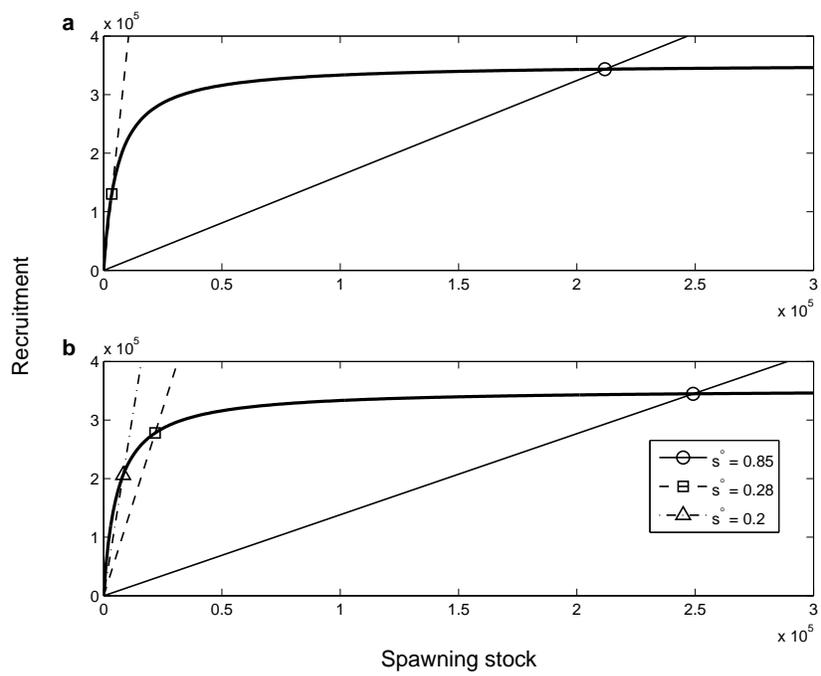}
\end{center} \caption{Graphical interpretation of the equilibrium
recruitment as the intersection of the smolt-adult curve and a line
through the origin with slope 1/(lifetime reproduction) for (a)
chinook and (b) coho salmon.}  \label{fig:coho-chinook-equilibria}
\end{figure}

Each of these models without random fluctuation has a unique positive
fixed point (equation (\ref{eq:fixed-point})).  The condition for a fixed point
can be illustrated in terms of the stock-recruitment function and a
straight line through the origin with slope equal to the inverse of
mean lifetime egg production, as shown in
Fig.~\ref{fig:coho-chinook-equilibria}
\citep{sissenwine:1987:alte}. The fixed point lies at the intersection
of these, allowing clear interpretation of the effects of long-term
changes in survival. As mean ocean survival declines, equilibrium egg
production declines, and eventually recruitment declines.  For the
very low survival rate for chinook salmon, the fixed point is zero
indicating survival is not adequate for persistence.  Because the
parameters for the stock-recruitment relationship and the estimated
survivals are for coho salmon, this has no implications for real
chinook salmon populations.

\begin{figure}
  \centering
  \includegraphics[width=0.5\textwidth]{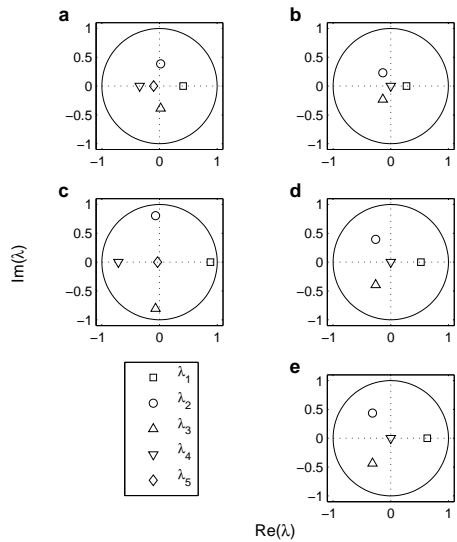}
  \caption{Eigenvalues of linearization matrix $\mathbf{J}$ in chinook and coho models for typical survival: (a) chinook, $s=0.85~\mathrm{yr}^{-1}$, (b) coho, $s=0.85~\mathrm{yr}^{-1}$; small survival: (c) chinook, $s=0.28~\mathrm{yr}^{-1}$, (d) coho, $s=0.28~\mathrm{yr}^{-1}$; and very small survival: (e) coho, $s=0.2~\mathrm{yr}^{-1}$.}
  \label{fig:coho-chinook-eigenvalues}
\end{figure}

 Plots of  the
eigenvalues of the linearization at the fixed point of each of these
models indicate the modes of variability, and how they change with survival (Figure~\ref{fig:coho-chinook-eigenvalues}).
Note that the chinook model has 5 eigenvalues, while the coho model has 4 eigenvalues because those are the maximum ages in each model population.  In both the chinook and coho models, the linearized population
dynamics respond most strongly to forcing in the
one-di\-men\-sion\-al mode corresponding to the positive real
eigenvalue, $\lambda_1$ (the dominant mode of variability).  Motion in
this mode occurs without oscillation, that is, at low frequencies. The next strongest mode is a resonance at period $n-1$ or a
little less (determined by the eigenvalues
$\lambda_2$ and $\lambda_3$, whose angle from the positive axis is
approximately $\pm \frac{2\pi}{n-1}$), indicating oscillations with a
period equal to one generation time.  In the chinook case (Figures 3a,c) there is
also a strong resonance at period two, corresponding to a negative
eigenvalue, and in all cases there is a weakly resonating negative
eigenvalue as well.

In both the chinook and the coho cases, the effect of decreasing the
mean survival is to move all eigenvalues outward toward the unit
circle, increasing the return time of all subsystems of interest and
thereby increasing the expected magnitude of the cumulative response
to variation over time.  This effect is especially strong in the
chinook model for ``small'' $s$
(Fig.~\ref{fig:coho-chinook-eigenvalues}c). This occurs because the
rate of change of recruitment with adult stock increases substantially
as the survival declines
(Fig.~\ref{fig:coho-chinook-equilibria}). Therefore, low equilibrium abundances in Figure~\ref{fig:coho-chinook-equilibria} correspond to large magnitude eigenvalues (and, thus, large responses to variation)in Figure~\ref{fig:coho-chinook-eigenvalues}.

The spread in ages of spawning has a relatively small specific influence on the
locations of the eigenvalues.  When $\sigma=0$, all fish spawn at age $n-1$, and the eigenvalues are
roots of a simple characteristic polynomial, $P(\lambda) =
\lambda(\lambda^{n-1}-s^{n-2}\frac{c^2}{\alpha})$. There is one zero
root and the others are equal in size and evenly spaced around zero
(not shown).  As $\sigma$ increases, the
dominant, positive real eigenvalue moves outward on the negative real axis, the others
move inward slightly and the complex eigenvalues rotate slightly, their polar angles
becoming slightly smaller in the chinook case and slightly larger in
the coho case.  
In all cases other than $\sigma=0$, the positive
eigenvalue is largest in magnitude (consistent with the
Perron-Frobenius theorem, since the entries of $\mathbf{J}$ are
nonnegative).  For our purposes, we conclude that the spread in
spawning age has very little impact on the model dynamics (too slight
to justify an illustration, in fact), and so our models do not
indicate that it is an important difference between chinook and coho
population dynamics.

\subsection{Mechanism of environmental forcing}

To illustrate the difference between population responses to
different mechanisms of environmental forcing we compare the cases
with (1) variability in survival in each ocean year, (2) variability
in survival at the age of ocean entry only and (3) variability in the
spawning age distribution.  \comment{For each of these cases we use frequency
responses (see Appendix) and time series to present an example using
one of the five models in Fig.~\ref{fig:coho-chinook-eigenvalues},
then describe the similar responses in the other four models.}
Random variability in survival appears to preferentially excite the geometrically decaying mode, while variability in the spawning age preferentially excites the cyclic mode.  This conclusion holds for all five of our model cases: the resonant
frequency is about $0.33~\mathrm{yr}^{-1}$ for the coho salmon, whose dominant age of spawning is three years, and about $0.25~\mathrm{yr}^{-1}$ for the chinook salmon, whose dominant age of spawning is four years
(not shown here, see next example).  The one exception is that for the chinook model with
$s^\circ=0.28~\mathrm{yr}^{-1}$, there is a strong low-frequency component with variation in
spawning age as well as the period-4 component.

\begin{figure}
  \begin{center}
    \includegraphics[width=0.9\textwidth]{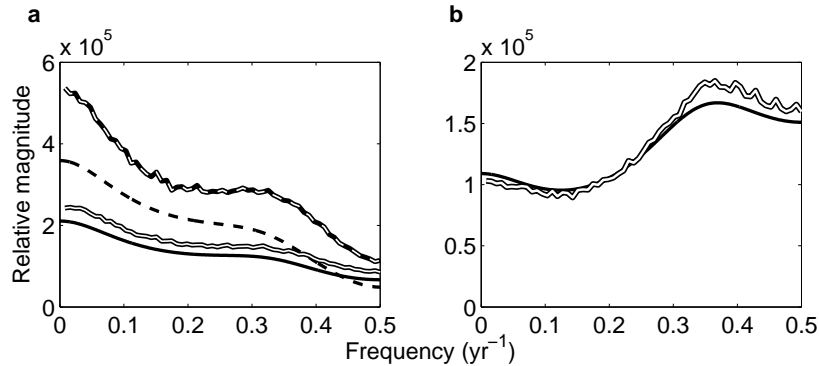}
\end{center}
\caption{Magnitude of frequency response of recruitment for the
  linearized model (thin lines) and nonlinear simulations (bordered lines) for environmental forcing of (a)
  early ocean survival rate (solid lines) and ocean survival at all ages (dashed lines), and (b) spawning age distribution for coho salmon with
  $s^\circ=0.28~\mathrm{yr}^{-1}$.
See the Appendix (equation~\ref{eqn:linear-frequency-response}) for derivation of the linearized frequency reponse.
Nonlinear frequency response curves are estimated by calculating the
average magnitude (the square root of mean power at each
frequency) of the fast Fourier transforms of 1000 simulation time series of length 128, and scaling the result by $\sqrt{128}~\sigma_E$, to obtain the
same units as the analytically-derived transfer function.}
\label{fig:transfer-coho-s_se-vs-dev}
\end{figure}

We illustrate this general result with the case for coho salmon with
$s^\circ=0.28~\mathrm{yr}^{-1}$ by
comparing the frequency responses to survival varying at all ages, 
for survival varying only at age 1, the presumed age of ocean entry, and for
variability in spawning age, then presenting examples of time series for each
case.  Both frequency responses to environmental variability in survival
(Fig.~\ref{fig:transfer-coho-s_se-vs-dev}a) decline from very low
frequency, leveling off slightly at a frequency just below that
corresponding to period 3 (the dominant age of spawning for coho
salmon), then decline for higher frequency. The simulations indicate
up to approximately 20 percent greater variability than the analytical model in both
survival cases.  The case with variable ocean survival at all ages
(Fig.~\ref{fig:transfer-coho-s_se-vs-dev}b) is skewed more toward
variability at low frequencies. The frequency response to
environmental variability in the spawning age distribution increases
from low frequency to a resonance at a frequency slightly greater than
that corresponding to period three, the dominant age of spawning, consistent with cohort resonance.

\begin{figure}
  \begin{center}
    \includegraphics[width=0.9\textwidth]{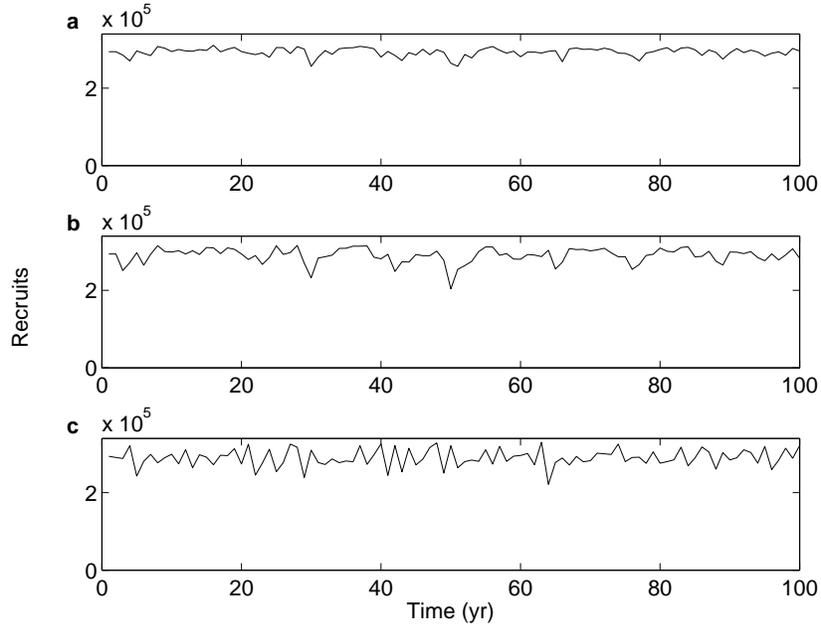}
\end{center}
\caption{Time series of recruitment in simulations with environmental forcing of (a)
  early ocean survival ($\sigma_E = 0.085$), (b) ocean survival at all ages ($\sigma_E = 0.085$), and (b) spawning age distribution
  ($\sigma_E = 0.2$) for coho with $s^\circ=0.28~\mathrm{yr}^{-1}$. Identical sequences of
  standard normal random variables were used as the forcing signal in
  order to allow a comparison of filtering by different demographic
  mechanisms.}
\label{fig:ts-coho-s_se-vs-dev}
\end{figure}

The time series for each case (Fig.~\ref{fig:ts-coho-s_se-vs-dev})
indicate a visually discernable difference between populations driven
by environmental variability in survival and environmental variability
in spawning age distribution. The population with environmental
forcing of spawning age (Fig.~\ref{fig:ts-coho-s_se-vs-dev}c) has
clear indications of cohort resonant behavior near period 2 yr to 4
yr, while the populations with environmental forcing of survival tend
toward 5-10 yr fluctuations with little variability.  The difference
in the magnitude of variability between the series with varying
survival is as expected from the difference in area under the two thin lines in
Fig.~\ref{fig:transfer-coho-s_se-vs-dev}a.

\subsection{Population observation}

To determine the effects of the type of observation on the time scales of variability we compared the results of observing recruitment to the results of observing total abundance for each case.  Generally, they all exhibited relatively more
2-to-5-year oscillation in the recruitment and more low-frequency
fluctuation in the total abundance.

\begin{figure}
  \begin{center}
    \includegraphics[width=0.9\textwidth]{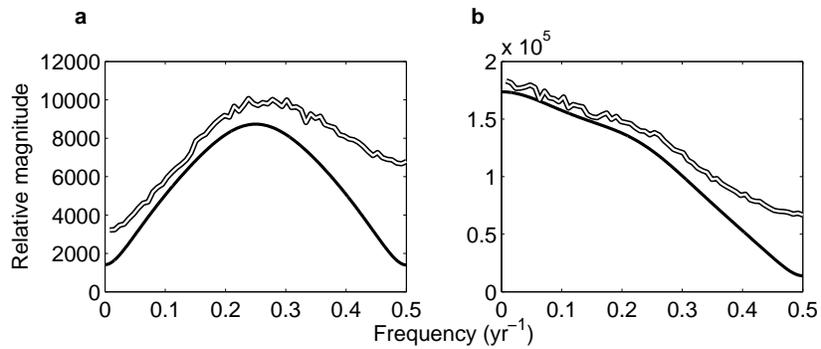}
\end{center}
\caption{Magnitude of frequency response of (a) recruitment and (b)
  total abundance for the linearized model (black line) and nonlinear
  simulations (bordered line; 1000 simulations of length 128) for
  environmental forcing of spawning age distribution for chinook salmon with
  $s = 0.85~\mathrm{yr}^{-1}$.}
\label{fig:transfer-chinook-R-vs-N-high-s}
\end{figure}

To illustrate the difference between different types of population
observations, we show the results of observing recruitment with
the results of observing total abundance, for chinook salmon with
typical survival ($s=0.85~\mathrm{yr}^{-1}$) and environmental variability in the spawning age
distribution (Fig.~\ref{fig:transfer-chinook-R-vs-N-high-s}).  
  The spectral
response of recruitment increases from low frequency to a peak at about $0.25~\mathrm{yr}^{-1}$, the frequency expected for chinook salmon with dominant
age of reproduction at 4 yr, then declines.  The spectral response of
total abundance, the sum of several cohorts, declines monotonically
from low frequency, showing only a hint of the resonance present in
recruitment.

\begin{figure}
  \begin{center}
    \includegraphics[width=0.9\textwidth]{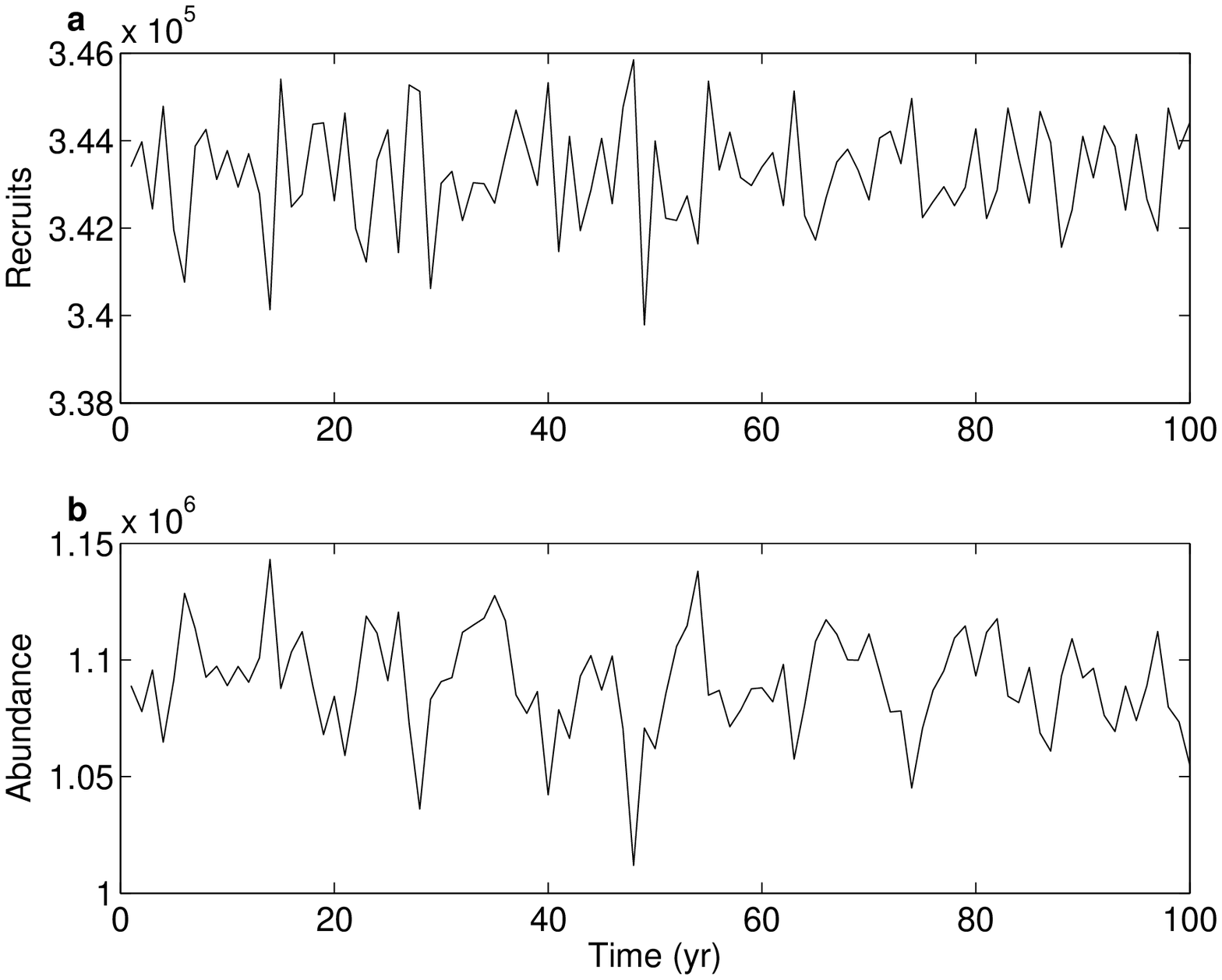}
\end{center}
\caption{Time series of (a) recruitment and (b) total abundance from a single
  simulation with environmental forcing of spawning age distribution
  ($\sigma_E = 0.2$) for chinook salmon with $s=0.85~\mathrm{yr}^{-1}$.}
\label{fig:ts-chinook-R-vs-N-high-s}
\end{figure}

The time series of these two cases
(Fig.~\ref{fig:ts-chinook-R-vs-N-high-s}) reflect these
characteristics.  The time series from the population with
observations of recruitment of a chinook salmon population with
environmental forcing of spawning age appears to have a preponderance
of variability on time scales of 2 yr to 5 yr, while the time series
from the observation of abundance appears to be a smoothed version of
that.

\comment{ 
\marginnote{To do: add a paragraph at the end of this section addressing whether this conclusion holds for all 5 cases.}
chinook small: yes - both red and blue, but more red, less blue in recruitment
chinook large: yes
coho very small: yes
coho small: yes
coho large: yes
}

\subsection{Long-term mean survival}

\begin{figure}
  \begin{center}
    \includegraphics[width=0.45\textwidth]{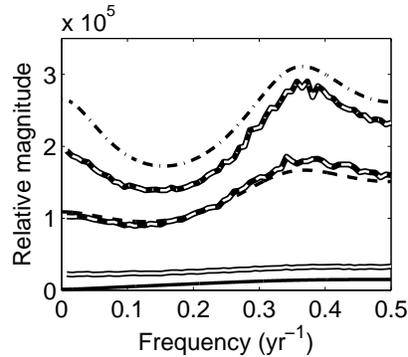}
\end{center}
\caption{Magnitude of frequency response of recruitment for the
  linearized model (thin lines) and nonlinear simulations (bordered
  lines; 1000 simulations of length 128 with $\sigma_E = 0.2$) for
  environmental forcing of spawning age distribution for coho with $s=0.85~\mathrm{yr}^{-1}$ (solid lines), $0.28~\mathrm{yr}^{-1}$ (dashed lines), and $0.2~\mathrm{yr}^{-1}$ (dash-dot
  lines).}
\label{fig:transfer-coho-dev-all-s}
\end{figure}

To illustrate the differences between populations operating at
different levels of long-term survival we compared the coho salmon
recruitment from a model with environmental variability in spawning
age distribution with each of the three survival levels, typical,
small and very small (Fig.~\ref{fig:transfer-coho-dev-all-s}). The shift to lower constant survivals could result from fishing or a shift in climate.  The
frequency responses with these three survivals have
similar shapes but differ substantially in scale.  The resonance at period
just less than 3 years is dominant because the variability is in
spawning age distribution.  Note the greater increase in sensitivity to low frequencies as the survival decreases to its lowest level.

\begin{figure}
  \begin{center}
    \includegraphics[width=0.9\textwidth]{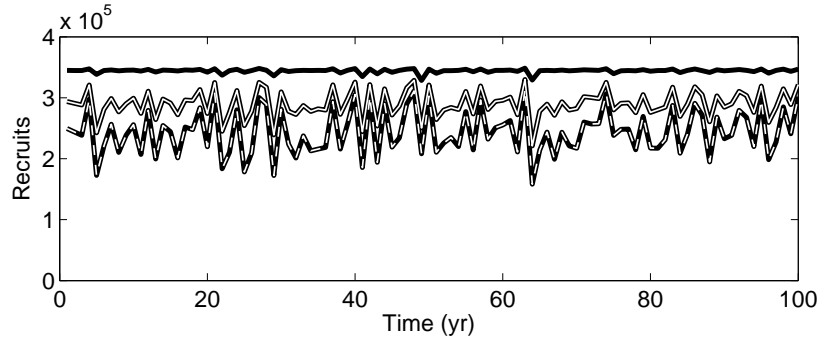}
\end{center}
\caption{Time series of recruitment in simulations with environmental
  forcing of spawning age distribution ($\sigma_E = 0.2$) for coho with
  $s=0.85~\mathrm{yr}^{-1}$ (solid black), $0.28~\mathrm{yr}^{-1}$ (white with black border), and
  $0.2~\mathrm{yr}^{-1}$ (dashed white on black). Identical sequences of standard
  normal random variables were used as the forcing signal in order to
  allow a comparison of filtering at different survival rates.}
\label{fig:ts-coho-dev-all-s}
\end{figure}

The time series of these three cases
(Fig.~\ref{fig:ts-coho-dev-all-s}) appear to have similar frequency
content, but different levels of variability as expected from Fig.~\ref{fig:transfer-coho-dev-all-s}.  
Importantly, they also underscore the fact that as survival declines
variability increases as in Fig.~\ref{fig:transfer-coho-dev-all-s}, as the equilibrium recruitment
declines.

\section{Discussion}

These analyses provide an understanding of how various population
characteristics shape the response of salmon populations to
environmental variability on various time scales.  Salmon population variability
does not simply follow variability in the environment as is commonly
assumed, but rather the observed response is shaped by three factors: (a)
the life history point of impact of the environment, (b) how the population is censused, and (c) pre-conditioning by long-term
changes in survival.  These considerations have important consequences for the
management of salmon populations and the anticipation of the effects
of large-scale environmental change.  The dynamics of salmon
responses to the environment are, of course, not unique, but rather
are closely related to those of other higher trophic level,
age-structured species \citep{mccann:2003:diff,bjornstad:2004:tren}.

The importance of cohort resonance to climate change is enhanced by the finding that its effects intensify with decreasing survival.  There is concern over the effects of climate change on fisheries, and how management should change to mitigate those changes \citep[\eg][]{perry:2010:sens}.  The basic resonance mode with period $T$ appears as the second and third eigenvalues in the linearized model, a conjugate pair.  It is well known in
linear age-structured models, as the echo effect
\citep{sykes:1969:ondi,caswell:2001:matr}, and was identified earlier
as a potential cause of observed cycles in some salmon populations
\citep{myers:1998:simp}.  However, increasing cohort resonance with fishing has not been mentioned as a factor in the variability in fished populations.  Knowing that high levels of fishing and specific time scales of environmental variability can increase overall population variability will be valuable in formulating management policies to maintain sustainability.  The additional variability increases the risk of population collapse in addition to that risk incurred by reducing mean abundance.   In addition, knowing that fish populations could be more sensitive to some time scales than others will aid in the explanation of changes in fisheries' variability if the time scales of forcing change with climate.  Potential intensification of cohort resonance through increased fishing also underscores the importance of Bj{\o}rnstad's \citeyearpar{bjornstad:2004:tren} warning regarding the low-frequency effect of cohort resonance, sensitivity to very low-frequency environmental variability.
Fishing increases the chances that populations, even when driven by white noise in
the environment, could generate very slowly changing signals that
could be mistaken for the effect of a slowly changing climate.  This
hard-to-detect long-timescale change may present a third increase in threat when
coupled with reduced abundance and increased variability due to fishing.  Note that this increased sensitivity at low frequencies was seen in coho salmon (Fig. 8) and also in chinook salmon with $s=0.28~\mathrm{yr}^{-1}$ (as noted at the end of section 4.2).

Differences in the dynamic responses of populations to temporal
variability at different points in the species's life history are not
commonly considered.  For salmon (and most other marine fish), while the effects of variability in survival are widely appreciated, the authors are unaware of any description of the effects of variability in spawning age distribution on population dynamics, such as on cohort resonance.  Here we found that the different mechanisms involving
variability in survival and variability in spawning age distribution
preferentially excited different fundamental modes of variability;
hence these differences are clearly important.  In addition to variability in
spawning age exciting a completely different mode than variability in
survival, responses to varying survival at ocean entry
and varying total survival also differed, with less sensitivity to low frequencies
resulting from variability in early ocean survival than from variability in total survival.  This further motivates efforts to discover the ages and locations at which variability in salmon marine survival occurs.
This may be related to the difference in population persistence between these two
survival mechanisms in earlier analysis \citep{botsford:2005:diff},
where population variability was greater when it occurred at the age
of return to the river for spawning, rather than the age of ocean
entry of smolts.  In that case this difference occurs because of the Law of Large Numbers, and
the fact that the logarithm of the number of spawners is the sum of several random
survivals when variability is at the age of entry, but only one random
survival when it is at the age of return.


In our models, varying age of spawning causes much more oscillation at the period of the
generation length and less low-frequency fluctuation than varying survival
does.  This might be because changes in central spawning age have a double
effect on recruitment, removing fish from one year's stock of potential spawners and adding
them to the next or previous year's potential spawning stock. This perturbation has a direct
effect on two cohorts' recruitment, which is echoed in subsequent
generations when those cohorts spawn. Fluctuations in survival, on the
other hand, whether they affect one cohort or many, cannot have this
double effect, with the likely consequence that the period-$T$ echo effect is not
as extreme.
   
The difference in frequency content between a catch (or abundance)
series and a recruitment series was known previously
\citep[\eg][]{botsford:1986:effe}, but there has been renewed
interest in it. That difference basically follows from the Law of
Large Numbers and the fact that when recruitment is the source of
temporal variability, abundance will have a lower coefficient of
variation because it is the sum of several recruitments, and hence can be expected to be a smoothed version of the recruitment signal, \ie to emphasize lower frequencies.  As
the value of constant harvest rate increases, the number of cohorts
summed diminishes, and hence higher frequencies are observed.  

The major difference between the population dynamics of coho salmon
and chinook salmon revealed here is that they would be most sensitive
to different time scales of environmental variability, as determined
by the difference between the dominant age of spawning in each.  Also, one
explanation proposed for the fact that coho salmon declined near 1980,
while chinook salmon did not, was that narrower width of the coho
spawning age distribution.  The effect of width of the spawning age
distribution we determined here was not strong enough to support this
hypothesis, leaving open the possibility that coho salmon were merely
following a decline in ocean survival which did not decline for
chinook salmon.  These theoretical results are not definitive
regarding specific stocks; rather they provide a context for further
detailed investigations of the hundreds of coho and chinook salmon
populations along the west coast of the U.S.

As with other structured-population models
\citep{nisbet:1982:mode,bjornstad:2004:tren,greenman:2003:ampli}, this
model's response to environmental fluctuation is predicted well by the
linearization at the deterministic model's fixed point.  Like Greenman
and Benton, we find amplification of noise as parameters vary --- in
our case, environmental noise is amplified more by the population
dynamics when survival is reduced.  In our case, however, this
amplification is not associated with a nearby bifurcation, only with
eigenvalues moving closer to the unit circle, a more general
phenomenon.  

As discussed above, we also observe major differences in frequency
responses among different kinds of environmental influences and
between different population measurements.  The analysis in our
Appendix explains these differences in terms of different
geometric relationships between the forcing vectors, the eigenspaces
corresponding to the eigenvalues of the linearization, and the
measurements.  The projection of the forcing vectors $\vec{H}$ into
each dynamical subspace determines how strongly the environmental noise
stimulates motion in that subsystem, and therefore how prominent
resonance at that frequency is in the population dynamics.  Similarly,
different observations --- total population and recruitment ---
include the different subsystems in different proportions, and so each
mode of population behavior is more visible in some observations than
others (see the Appendix for the mathematical treatment of these
ideas).  This analysis, together with the diversity in frequency
responses we see in our models, points to the importance of
understanding the relationship between the eigenspace structure of the
linearization and the relative importances of the resonant frequencies
associated with each eigenspace.  As an analytical technique, this
approach may be useful in multispecies ecological models as well as in
structured population models.

The results obtained here provide the means both to begin to explain current differences in responses to the environment by the same species at different locations, and to project differences in future responses on the basis of projected changes in time scales of variability in the environment.  Current differences in responses by populations of the same species are typically presumed to imply a difference in environmental forcing, but they may be due to differences in life histories or differences in preconditioning because they are fished at different intensities.  Future changes in time scales of variability are expected on the basis of paleological records \citep[\eg past changes in the time scales of variability of El Ni\~{n}o events as observed in corals and other media,][]{Jones:2009:high} or predictions from global climate models \citep{Timmermann:1999:incr}.

\section*{Acknowledgments}
This research is part of the GLOBEC Northeast Pacific program and was funded by  National Science Foundation grants NSF OCE0003254 and NSF OCE0815293.  Any
opinions, findings, and conclusions or recommendations expressed in this
material are those of the author(s) and do not necessarily reflect the views
of the National Science Foundation.

\bibliographystyle{elsarticle-harv}
\bibliography{Worden_et_al_review}

\newpage
\appendix
\section{General Mathematical Results}

The linearization of a stochastic map
\begin{equation}
\vec{x}(t)=F(\vec{x}(t-1), \xi(t), \xi(t-1), \ldots, \xi(t-L)),
 \quad \vec{x}(t)\in\mathbb{R}^n, \quad \xi(t) \in \mathbb{R},
\end{equation}
where (without loss of generality) $\mathrm{E}(\xi(t))=0$, is
\begin{equation}
\vec{y}(t) \approx \mathbf{J}\vec{y}(t-1) + \sum_{l=0}^L \vec{H}(l)\xi(t-l)
\end{equation}
where $\vec{x}(t)=\tilde{\vec{x}} +\vec{y}(t)$, $\tilde{\vec{x}}$ is a
fixed point of $F(\vec{x},0)$, 
$\mathbf{J}=\left(\frac{\partial F_i}{\partial x_j}\right)|_{(\tilde{\vec{x}},0)}$ is the Jacobian
matrix of $F$ at $\vec{x}=\tilde{\vec{x}}$ and $\xi=0$,
$\vec{H}(l)=\frac{\partial F}{\partial{\xi(t-l)}}(\tilde{\vec{x}},0)$
is a vector expressing the dependence of $F$ on the noise terms, and $L$ is the maximum time lag at which
stochastic perturbations affect $F$ directly. Let $\vec{v}_i$ and
$\vec{u}_i$ be the left and right eigenvectors of
$\mathbf{J}$, respectively, and $\lambda_i$ its eigenvalues, so that
$\vec{v}_i \mathbf{J}=\lambda_i \vec{v}_i$ and
$\mathbf{J}\vec{u}_i=\lambda_i \vec{u}_i$. In this paper we only
consider matrices that have all distinct eigenvalues. 

We change to the natural coordinate system of
$\mathbf{J}$: Let
\begin{equation}
\mathbf{U}=(\vec{u}_1, \dots, \vec{u}_n), \;
 \mathbf{V}=\left(\begin{array}{c} \vec{v}_1\\ \vdots \\ \vec{v}_n \end{array} \right), \;
 \mathbf{\Lambda} = \left( \begin{array}{ccccc}
  \lambda_1 & 0 & \cdots &   & 0 \\
  0         &   &        &   &   \\
  \vdots    &   & \ddots &   & \vdots \\
            &   &        &   & 0 \\
  0         &   & \cdots & 0 & \lambda_n \end{array} \right),
\end{equation}
with
$\mathbf{J}=\mathbf{U}\mathbf{\Lambda}\mathbf{V}=\sum_{i}\lambda_{i}\vec{u}_{i}\vec{v}_{i}$
and $\mathbf{UV}=\mathbf{VU}=\mathbf{I}$. Then for $\vec{w}=\mathbf{V}\vec{y}$,
\begin{equation} \begin{split} \label{b}
\vec{w}(t)=\mathbf{V}\vec{y}(t) &=\mathbf{\Lambda} \mathbf{V} \vec{y}(t-1) + \sum_{l=0}^L \vec{G}(l) \xi(t-l) \\
 &= \mathbf{\Lambda} \vec{w}(t-1) + \sum_{l=0}^L \vec{G}(l)\xi(t-l)
\end{split} \end{equation}
with $\vec{G}(l)=\mathbf{V}\vec{H}(l)$. 
Using the terms of this transformed vector $\vec{w}$, the vector of
deviations from equilibrium can be written as a sum of eigenvectors,
$\vec{y} = \sum_i w_i \vec{u}_i$.
Since $\mathbf{\Lambda}$ is diagonal, the
dynamics of each $w_i (t)$ is uncoupled from the others:
\begin{equation}
w_i (t) = \lambda_i w_i (t-1) + \sum_{l=0}^L g_i(l) \xi(t-l)
\end{equation}
where $g_i(l) = \vec{v}_i \vec{H}(l)$ is the $i$th entry of
$\vec{G}(l)$.  Thus, $\vec{w}(t)$ is the state of the linear
system decomposed into its independent subsystems, and
the vectors $\vec{G}(l)$ represent the stochastic forcing resolved
into the decomposed coordinate system.

\subsection*{Forcing and measurement}

The entries of the transformed vector $\vec{G}$ reveal how strongly
the environmental forcing acts on each subsystem of the linearized
system.  Different kinds of forcing (i.e. survival, age of maturation)
are distributed differently among the different subsystems,
characterized by the projection of the forcing into each eigenspace,
$g_i(l)=\vec{v}_i\vec{H}(l)$ for each $i$ and $l$.  If
$g_i(l)=0$ for all $l$, there is no fluctuation in the subspace
containing eigenvector $\vec{u}_i$, that is, no fluctuation in $w_i$.
In general, the more $\vec{H}$ is aligned with
certain eigenvectors $\vec{u}_i$, the more the fluctuations caused by
the forcing signal will be concentrated in those subsystems.

Similarly, a particular measurement generally observes some subsystems
more than others.  Assume we are observing the population via a scalar
measurement, whether annual total population, recruitment or catch,
represented as $M(t) = Q(\vec{x}(t))$.  Let us assume that
$Q(\vec{x})$ is differentiable at the fixed point $\tilde{\vec{x}}$.
In the weak-noise limit this quantity also can be described by a
linear approximation,
\begin{equation} \begin{split}
Q(\vec{x}(t)) &= Q(\tilde{\vec{x}})
            + \sum_i \frac{\partial Q}{\partial x_i}(\tilde{\vec{x}})\,y_i(t)
            + \BigO(|y|^2) \\
        &= Q(\tilde{\vec{x}}) + \sum_i q_i\,y_i(t) + \BigO(|y|^2).
\end{split} \end{equation}
The measurement $M(t)$ is approximated by the linear quantity
$Q(\tilde{\vec{x}})+\sum_iq_i\,y_i(t)$, which differs by only a
constant from $\sum_iq_i\,x_i(t)$, and fluctuation in either quantity
has the same characteristics as fluctuation in
$\sum_iq_i\,y_i(t)$.  We can represent the linearized measurement as a
vector product $\vec{q}\,\vec{y}(t)$, using a row vector
$\vec{q}=(q_1,\ \dots,\ q_n)$.  Changing to the natural coordinates of
$\mathbf{J}$,
\begin{equation} \begin{split}
\vec{q}\,\vec{y}(t)
 &= \vec{q} \sum_i w_i(t)\,\vec{u}_i 
 \;\;=\;\; \sum_i \vec{q}\,\vec{u}_i\,w_i(t)
 \;\;=\;\; \sum_i m_i\,w_i(t).
\end{split} \end{equation}
Then $m_i = \vec{q}\,\vec{u}_i$ expresses how strongly the motion
in the subsystem containing $w_i$ is reflected in the measurement.

\subsection*{Frequency Analysis} 

For the frequency analysis, take $\vec{w}(t)=\vec{0}$ and $\xi(t)=0$ for all $t<0$.
Then we may write the $Z$-transform \citep{Elaydi:1999}:
\begin{equation} \begin{split}
\hat{w}(z) &\equiv \sum_{k=0}^\infty \vec{w}(k)\,z^{-k} \\
 &= \sum_{k=0}^\infty \left(\mathbf{\Lambda}\,\vec{w}(k-1) + \sum_{l=0}^L\vec{G}(l)\,\xi(k-l)\right)\,z^{-k} \\
&= \mathbf{\Lambda}\underbrace{\sum_{k=0}^\infty \vec{w}(k-1)\,z^{-k}} +
   \sum_{l=0}^L\vec{G}(l)\underbrace{\sum_{k=0}^\infty \xi(k-l)\,z^{-k}}.
\end{split} \end{equation}
The bracketed expressions are themselves $Z$-transforms of shifted
sequences of $\vec{w}$ and $\xi$, so that by the shift property
of the $Z$-transform \citep{Elaydi:1999},
\begin{equation} \begin{split}
  \hat{w}(z) = \mathbf{\Lambda}\,z^{-1}\hat{w}(z) + \sum_{l=0}^L\vec{G}(l)\,z^{-l}\hat{\xi}(z)\
\end{split} \end{equation}
or
\begin{equation} \label{eqn:linear-frequency-response}
\hat{w}(z) = (1-\mathbf{\Lambda}\,z^{-1})^{-1} \sum_{l=0}^L\vec{G}(l)\,z^{-l}\hat{\xi}(z).
\end{equation}
$(1- \mathbf{\Lambda} z^{-1})^{-1}$ is a diagonal matrix with entries
$\frac{1}{1- \lambda_i z^{-1}} = \frac{z}{z-\lambda_i}$.  Consequently
the $Z$-transformed deviation vector,
\begin{equation}
\hat{y}(z)=\mathbf{U}\hat{w}(z)=\sum_{i=1}^n\left(\vec{u}_i\,\frac{z}{z-\lambda_i}\,\sum_{l=0}^L g_i(l)\,z^{-l}\hat{\xi}(z)\right),
\end{equation}
is a linear sum of rational functions of $z$, with peaks tending to be near the eigenvalues of $\mathbf{J}$.

Similarly, in the frequency analysis of a measurement $M(t)$,
\begin{equation} \begin{split}
\hat{M}(z)
 &= \sum_{k=0}^\infty M(k)z^{-k} \\
 &\approx \sum_k (Q(\tilde{\vec{x}}) + \sum_i m_i\,w_i(k))\,z^{-k} \\
 &= \sum_k Q(\tilde{\vec{x}})\,z^{-k} + \sum_i m_i\,\hat{w}(z).
\end{split} \end{equation}
It is appropriate to discard the constant part of $M(t)$ since we are
concerned with year to year variation:
\begin{equation} \begin{split}
\sum_i m_i\,\hat{w}(z)
 &= \sum_i \left[m_i\sum_{l=0}^Lg_i(l)\,\frac{z}{z-\lambda_i}\,z^{-l}\hat{\xi}(z)\right] \\
 &= T_M(z)\,\hat{\xi}(z).
\end{split} \end{equation}
The frequency response $T_M(z)$ of the measurement $M(t)$ is a
weighted sum of the frequency responses for each subsystem,
$T_i(z)=\frac{z}{z-\lambda_i}$, each weighted both by the
strength of environmental forcing in that subsystem at each lag,
$g_i(l)$, and by the ``emphasis'' of the subsystem in the
measurement, $m_i$.

\end{document}